\documentclass[10pt,journal,compsoc]{IEEEtran}
\usepackage{graphicx}
\usepackage{caption}
\usepackage{subcaption}
\usepackage[utf8]{inputenc}
\usepackage{authblk}
\usepackage{todonotes}
\makeatletter
\let\oldtodo\todo
\renewcommand{\todo}[1]{\oldtodo[inline]{#1}}
\makeatother

\makeatletter
\usepackage{amsmath}
\usepackage{amsfonts}
\usepackage{amssymb}
\usepackage{hyperref}
\usepackage{cleveref}
\usepackage{comment}
\usepackage[ruled,vlined]{algorithm2e}

\newcommand{\ddKS}{ddKS}

\title{Accelerated Computation of a High Dimensional Kolmogorov-Smirnov Distance}
\author[1]{Alex Hagen}
\author[1]{Shane Jackson}
\author[2]{James Kahn}
\author[1]{Jan Strube}
\author[2]{Isabel Haide}
\author[1]{Karl Pazdernik}
\author[1]{Connor Hainje}
\affil[1]{Pacific Northwest National Laboratory, Richland, WA, USA}
\affil[2]{Karlsruhe Institute of Technology, Karlsruhe, Germany}
\date{\today{}}

\begin{document}

\maketitle

\begin{abstract}

Statistical testing is widespread and critical for a variety of scientific disciplines. The advent of machine learning and the increase of computing power has increased the interest in the analysis and statistical testing of multidimensional data.  We extend the powerful Kolmogorov-Smirnov two sample test to a high dimensional form in a similar manner to Fasano \cite{Fasano1987}.  We call our result the $d$-dimensional Kolmogorov-Smirnov test (\ddKS{}) and provide three novel contributions therewith: we develop an analytical equation for the significance of a given \ddKS{} score, we provide an algorithm for computation of \ddKS{} on modern computing hardware that is of constant time complexity for small sample sizes and dimensions, and we provide two approximate calculations of \ddKS{}: one that reduces the time complexity to linear at larger sample sizes, and another that reduces the time complexity to linear with increasing dimension.  We perform power analysis of \ddKS{} and its approximations on a corpus of datasets and compare to other common high dimensional two sample tests and distances: Hotelling's T$^2$ test and Kullback-Leibler divergence.  Our \ddKS{} test performs well for all datasets, dimensions, and sizes tests, whereas the other tests and distances fail to reject the null hypothesis on at least one dataset.  We therefore conclude that \ddKS{} is a powerful multidimensional two sample test for general use, and can be calculated in a fast and efficient manner using our parallel or approximate methods. Open source code for all methods accompanies this work\footnote{Open source implementations of all methods described in this work are located at \url{https://github.com/pnnl/ddks}.}.

\end{abstract}

\section{Introduction}

In the physical and data sciences, one-dimensional test statistics are commonly used to test whether two samples originated from the same distribution.  The one-dimensional Kolmogorov-Smirnov (OnedKS)\footnote{The Kolmogorov-Smirnov test is often called simply the KS test, we will refer to it as the OnedKS to differentiate from higher dimensional tests} test~\cite{Kolmogorov1933, Smirnov1948} is an easy to compute and nonparametric test statistic which compares the cumulative distribution functions (CDF) of two probability distributions. It is one of the most useful two-sample tests, as it is receptive to differences in both location, shape and spread of the CDFs. Indeed, the OnedKS test typically outperforms other widely used tests, such as the $\chi^2$-test \cite{Mitchell1971}, on smaller sample sizes; studies have shown it to be a powerful goodness of fit test akin to the Anderson-Darling \cite{razali2011power} or Cramer von Mises tests. While powerful in one-dimensional cases, these test statistics neglect any covariances between distributions, and, as such, are of limited discerning capabilities when working in higher dimensions.

Comparing multidimensional distributions is important in many fields, including climate science \cite{DeMichele2007}, astronomy \cite{Lotz2004}, social sciences \cite{Bharathi2021}, or, famously, the quality control of weapons \cite{Hotelling1947}.  More recently, machine learning and other data-driven methods have made high dimensional data more common and required the use of multidimensional hypothesis testing, such as in \cite{Zhu2021}.   The literature on divergences and distances between two high-dimensional distributions is robust. Common distribution distances are the Earthmover's (or Wasserstein) distance \cite{rubner1998metric}, and the Kullback-Leibler divergence \cite{Joyce2011}.  Most commonly used test statistics either make assumptions about the underlying data, e.g assuming an underlying distribution such as Hotelling's T-test \cite{Hotelling1931}, or are computationally expensive \cite{Fasano1987}.   

This paper addresses the two-sample problem of comparing data drawn from two probability distributions in higher dimensions through the multidimensional Kolmogorov-Smirnov test.  Generalizing the OnedKS test to higher dimensions while retaining the properties of the one-dimensional case has been done, e.g. by Fasano et. Al \cite{Fasano1987}. While statistically efficient, this method is computationally expensive, scaling as $\mathcal{O}(2^{d}N^2)$.  A time complexity comparison of multiple methods for 2 dimensional Kolmogorov-Smirnov test statistic calculations can be found in a review by Lopes et. Al \cite{lopes2008computationally}.  Methods for reducing the dimensional complexity exist; however they rely on calculating highest probability density (HPD) regions \cite{harrison2015validation} or minimum volume (MV) sets \cite{glazer2012learning, polonik1999concentration}. Both methods ostensibly reduce time complexity by reframing the problem in terms of a singular parameter (the amount of the input volume associated with some probability $\alpha$). This reframing belies the difficulty in calculating such regions, especially in high dimensions.  Some form of approximation, such as using machine learning approximators to learn level sets for estimation of the MV sets, is used to make the problem tractable. Separately, some methods learn only local information about the underlying distribution \cite{Bendahan2017} as a way to reduce computational cost.

We instead choose the path of making the exact Kolmogorov-Smirnov test computationally tractable, and while we approximate the statistical distance, we apply no bias to the underlying distributions as the MV methods have done. One of the novel contributions of this work are accelerated methods of calculating the $d$-dimensional Kolmogorov-Smirnov \ddKS{} test as defined by Fasano et al. We present three accelerated methods, the first method trading time complexity with memory complexity while calculating \ddKS{} directly. The second and third method instead approximate \ddKS{} using voxelization- and sorting-based method, respectively, thus enabling a tradeoff between speed and statistical efficiency. In addition, an analytic calculation of the significance of a \ddKS{} statistic is derived.

The remainder of this paper is structured as follows. In Section \ref{sec:ddks} the \ddKS{} test statistic is presented, Section \ref{subsec:formalism} explaining the mathematical formalism of the calculation and Section \ref{subsec:acccomp} describing the computation of the three accelerated methods. The significance calculation is derived in Section \ref{sec:sig}, while Section \ref{sec:experiments} demonstrates the advantages of \ddKS{} in comparison to other widely used test statistics. We show that the \ddKS{} methods perform well on small sample sizes as well as on small distribution differences.

\section{ddKS Test Statistic}
\label{sec:ddks}

The $d$-dimensional extension of the Kolmogorov-Smirnov (\ddKS{}) test statistic \cite{Fasano1987}, is calculated similarly to the one-dimensional Kolmogorov-Smirnov test statistic: the \ddKS{} test statistic is the maximum of the differences between the cumulative distribution function (CDF) and survival function (SF) of two samples. The simplicity of this definition belies the difficulty in constructing the CDF and SF in dimensions higher than one. We lay out the formal definition of \ddKS{} in high dimensions in the next section.

\subsection{Formalism}
\label{subsec:formalism}
We begin with two finite sets of samples in $\mathbb{R}^d$ which we denote as $P$ and $T$ \footnote{$P$ and $T$ may be sampled from any distribution and in any order; however, to align with machine learning contexts, we refer to these as the ``predicted'' and ``true'' distributions, respectively.}.  We first assume that the Cartesian coordinate system is an appropriate basis for the samples provided.  Then, we can construct $2^{d}$ cumulative density estimates, where the cumulative density is defined as the number of points existing in an orthant relative to a given point.  This is equivalent in two dimensions to counting the number of points in the top-right, bottom-right, bottom-left, and top-left quadrants relative to a chosen location.  Given these two cumulative density estimates $C_{P}\left(\vec{x}\right)$ and $C_{T}\left(\vec{x}\right)$, the \ddKS{} test statistic is simply the maximum absolute difference between the two.
\begin{equation}\label{eq:ddks_definition}
    D = \max\left[ \left| C_{P}\left(\vec{x}\right) - C_{T}\left(\vec{x}\right) \right| \right]
\end{equation}
We note that following this definition, ddKS is a metric. The relevant proofs are given in Appendix \ref{sec:proofs}.

In practice, calculation of the maximum absolute differences between two functions defined over all space is impractical.  Several simplifications make the calculation straightforward and readily implemented on modern computing hardware.

Without applying bias, we cannot model any change in  $C_{P}$ and $C_{T}$ except at a point from the respective sample.  Therefore, we assume that $C$ only needs to be evaluated at each point in $P$ and $T$. We can now represent $C_{P}$ and $C_{T}$ not as functions, but as tensors of shape $\mathbf{C} \in \mathbb{R}^{t \times 2^d}$ where $t$ is the number of evaluation locations, naming these $\mathbf{C}_{P, x}$ and $\mathbf{C}_{T, x}$ with $x$ designating the set of evaluation locations. We note that the cumulative density function of $P$ evaluated at the locations of the points in $T$ is not guaranteed to be equal to the cumulative density function of $P$ evaluated at the locations of the points in $P$. This provides ambiguity as to which sample to use for evaluation locations\footnote{This ambiguity is very similar to the difference between extrinsic and intrinsic Kullback-Leibler Divergence.}. To alleviate this ambiguity, we define \ddKS{} as the maximum absolute differences between two cumulative distribution functions, evaluated at all points in both samples.
\begin{equation}\label{eq:ddks_tensor_definition}
    D = \max\left| \left(\mathbf{C}_{P,P} - \mathbf{C}_{T,P}\right)^\frown  \left(\mathbf{C}_{P,T} - \mathbf{C}_{T,T}\right) \right|
\end{equation}
where $^\frown$ denotes concatenation.

Calculation of equation \ref{eq:ddks_tensor_definition} is now implementable on modern computing hardware.  We built a loop based implementation in the following way: for each set of testing points in $P$ and $T$, we iterate through each test point.  At each test point, we define each orthant by finding all permutations of greater than or less than in each dimension, and determine the number of points from each sample that exist in that orthant.  We save the orthant membership of each class for each set of testing points, and concatenate the two membership matrices.  Then, we find the maximum absolute difference between the orthant membership of each class given the same set of testing points, which is an unbiased result to equation \ref{eq:ddks_tensor_definition}.

Unfortunately, the described loop based computation exhibits high computational complexity $\mathcal{O}\left(2^{d}N^{2}\right)$ where $N$ is the size of the combined sample set ($P$ and $T$). For many cases, this complexity is prohibitive; below we describe accelerated methods to alleviate the $N^2$ scaling, or separately the $2^{d}$ scaling.  

\subsection{Accelerated Computations}
\label{subsec:acccomp}
We have developed several accelerated methods of computing or approximating \ddKS{}.  The first of these methods directly computes \ddKS{}, but trades time complexity for memory complexity and is implemented in a tensor framework (\texttt{pytorch} \cite{Paszke2019}) for parallel computation on CPU or GPU.  The second and third approximate \ddKS{} using spatial indexing, grid-based and M-Tree inspired, respectively, to compute the orthant membership.  These enable tradeoffs between speed and statistical efficiency. Notably, the tensor based method computes the exact \ddKS{} test statistic with $\mathcal{O}\left(1\right)$ time complexity for $d$ and $N$ small enough such that the number of cores available, whether on CPU or GPU, is not exhausted. The voxel based method approximates \ddKS{} in $\mathcal{O}\left(2^{d}Nk\right)$ (where $k$ is the number of voxels) and so can be used for larger samples. The radial based method approximates \ddKS{} in $\mathcal{O}\left(\left(d+1\right)N\log N \right)$, thus providing a good method for larger dimensions. The time complexity of our accelerated methods, and some other methods for comparison, are shown in Figure~\ref{fig:timing}, tested on a single core CPU, a multiple core CPU, and GPU where possible.  The $\mathcal{O}\left(1\right)$ behavior is shown by the relatively constant time at the left side of every line for those methods that use tensor primitive computation (\ddKS{}, OnedKS, and KLDiv).  Hotelling's T$^2$ test exhibits very low computational complexity but is not as powerful as the other tests, as shown in Section~\ref{sec:results}. The subsequent sections describe the accelerated implementations of \ddKS{} which lead to the time complexity shown in Figure~\ref{fig:timing}.

\subsubsection{Tensor primitive based computation}
\label{subsec:tensor}
To enable parallel computation by using tensor primitives from the \texttt{pytorch} library, we implement orthant membership computations in single tensor computations.  First, we create two tensors from $P$ and $T$, expanding each along a new axis, copying all elements in the first two axes along the new axes, such that each is of size of the $P$ ($T$) tensor is $N_{P(T)} \times d \times N_{T(P)}$ where $N_x$ is the number of samples in the corresponding set. We call these new tensors $\mathbf{P}$ and $\mathbf{T}$.  We construct test points for the cumulative density function calculation by cloning $\mathbf{P}$ and $\mathbf{T}$ and permuting their first and third axes into tensors $\mathbf{S}_{P}$ and $\mathbf{S}_{T}$. By this process, $\mathbf{P}\left[i, j, k\right]$ is the $j$th dimensional element of the $i$th point, and $\mathbf{S}_{P}\left[i, j, k\right]$ is the $j$th dimensional element of the $k$th point.  The same holds true for $\mathbf{T}$ and $\mathbf{S}_{T}$.

We begin operations to calculate orthant membership.  First, we calculate the relationship of each point to test points in $\mathbf{S}_{P}$ and $\mathbf{S}_{T}$. This is 
\begin{align} \label{eq:g}
    \mathbf{G}_{P,P} &= \mathbf{P} \geq \mathbf{S}_{P}
    \quad{}
    \mathbf{G}_{T,P} = \mathbf{T} \geq \mathbf{S}_{P} \nonumber \\
    \mathbf{G}_{P,T} &= \mathbf{P} \geq \mathbf{S}_{T}
    \quad{}
    \mathbf{G}_{T,T} = \mathbf{T} \geq \mathbf{S}_{T}
\end{align}
where $\geq$ is the element-wise greater than or equal to operator, and each element in $\mathbf{G}$ is then cast to a floating point decimal.  With the $\mathbf{G}$ comparisons constructed, we then construct our orthant membership tensor by summing those points falling between partition hyperplanes.  In $d$ dimensions, there are $2^{d}$ partitions possible, so the orthant membership tensor $\mathbf{M}$ has size $N \times 2^{d}$. To separate points into orthants, we define a square wave function, varying between $0$ and $1$ with frequency $f$
\begin{equation}
    Sq\left(x, f\right) = \frac{\left(-1\right)^{\left\lfloor 2fx\right\rfloor} + 1}{2}
\end{equation}
Each element in the membership tensor is defined as
\begin{equation}
    \mathbf{M}_{X,Y}\left[i, j\right] = \sum_{k=1}^{N} \prod_{m=1}^{d} Sq\left(\mathbf{G}_{X,Y}\left[i, m, k\right], 2^{-m-2}\right)
    \label{eq::membership}
\end{equation}

The test statistic is defined as the largest difference between the orthant memberships of $\mathbf{P}$ and $\mathbf{T}$.  With $\mathbf{M}$ constructed, we calculate the differences between $\mathbf{P}$ and $\mathbf{T}$ centered at the points in $\mathbf{S}_{P}$ and $\mathbf{S}_{T}$ and find the absolute maximum of the union of these two sets as in~\ref{eq:ddks_tensor_definition}.
\begin{equation} \label{eq:d1}
D = \max\left[ \left| \left(\mathbf{M}_{P,P} - \mathbf{M}_{T,P} \right) ^\frown \left( \mathbf{M}_{P,T} - \mathbf{M}_{T,T}\right)\right| \right]
\end{equation}

\begin{figure*}
    \centering
    \includegraphics[width=\textwidth]{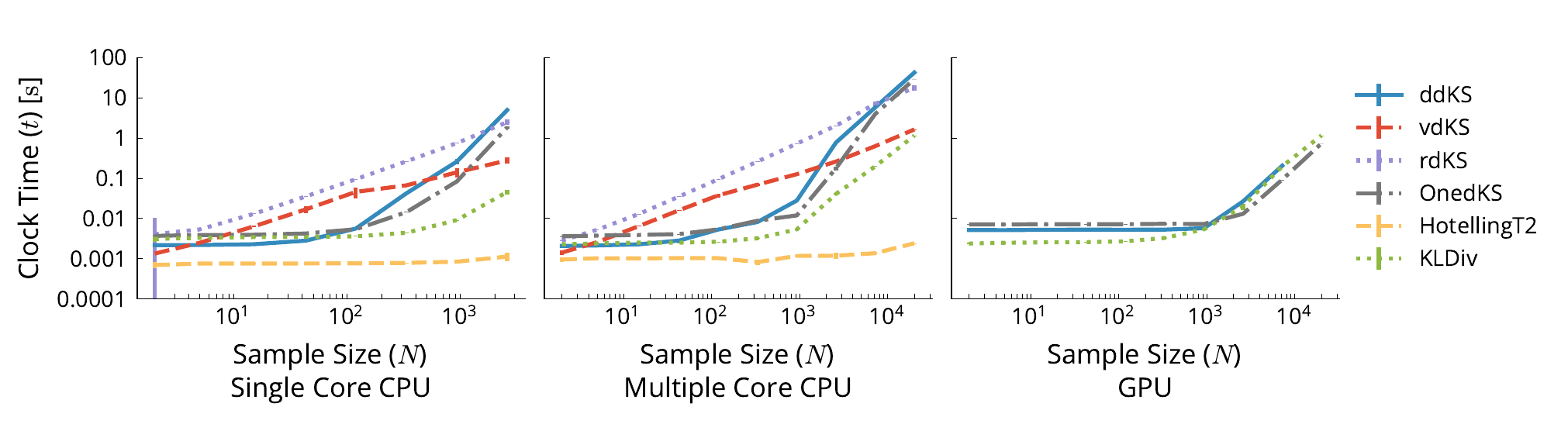}
    \caption{
    Time to compute a single test statistic for \ddKS{} accelerated methods and selected other methods from the literature.  A lower time for computation is better, however there is a tradeoff between time complexity as visualized here, and statistical power as visualized in Section~\ref{sec:results}.
    }
    \label{fig:timing}
\end{figure*}

This direct calculation method utilizes higher memory complexity than the loop based computation, but uses the implicit parallelization in \texttt{pytorch}'s (which exposes the threading and vectorization of Intel's Math Kernel Library) tensor primitive operations.  Tests show the computational complexity of this tensor based method to be $\mathcal{O}\left(1\right)$ until the processing unit's attached memory and core count is exhausted, whichever is first (see Figure~\ref{fig:timing} for one example of the number of samples at which this occurs).  For modern Graphics Processing Units (GPUs), this can extend up to $10,000$ points per sample in three dimensions.  Thus, this tensor based method can quickly calculate $D$ on modern data science workstations or GPU enabled accelerator.

\begin{figure}
    \centering
    \begin{subfigure}[b]{\columnwidth}
        \centering
        \includegraphics[width=0.7\textwidth]{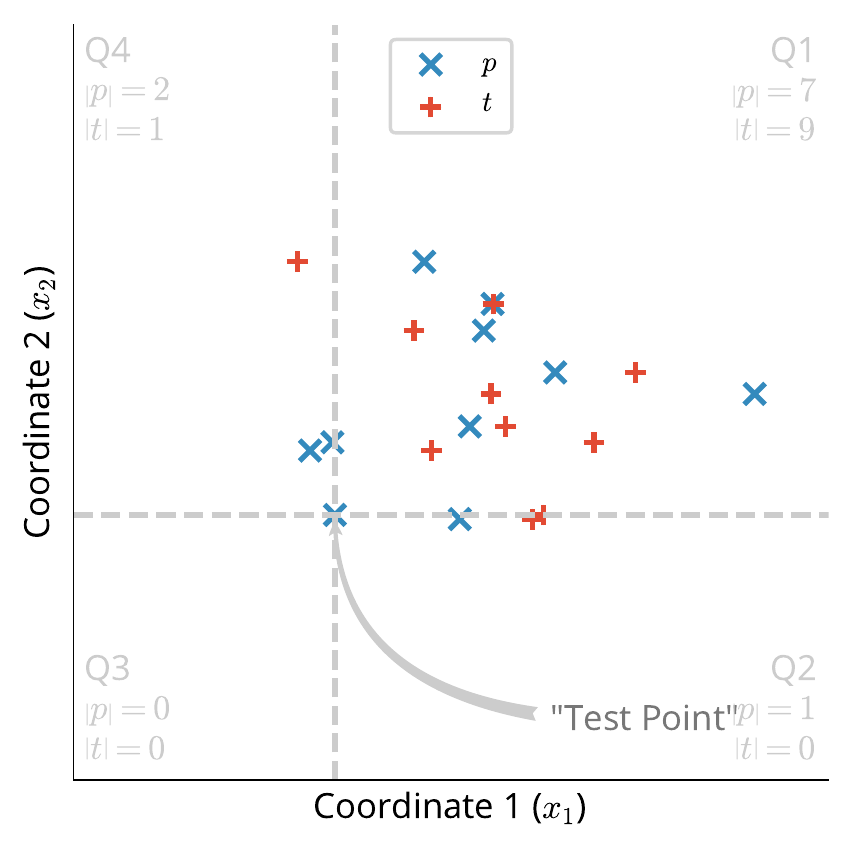}
        \caption{ddKS compares points to a test point along the basis vectors of the space, assigning membership to given quadrants based on their relationship to that test point. The normalized maximum difference in quadrant membership when calculated over all test points is the \ddKS{} statistic.}
        \label{fig:ddks_schematic}
    \end{subfigure}
    \begin{subfigure}[b]{\columnwidth}
        \centering
        \includegraphics[width=0.7\textwidth]{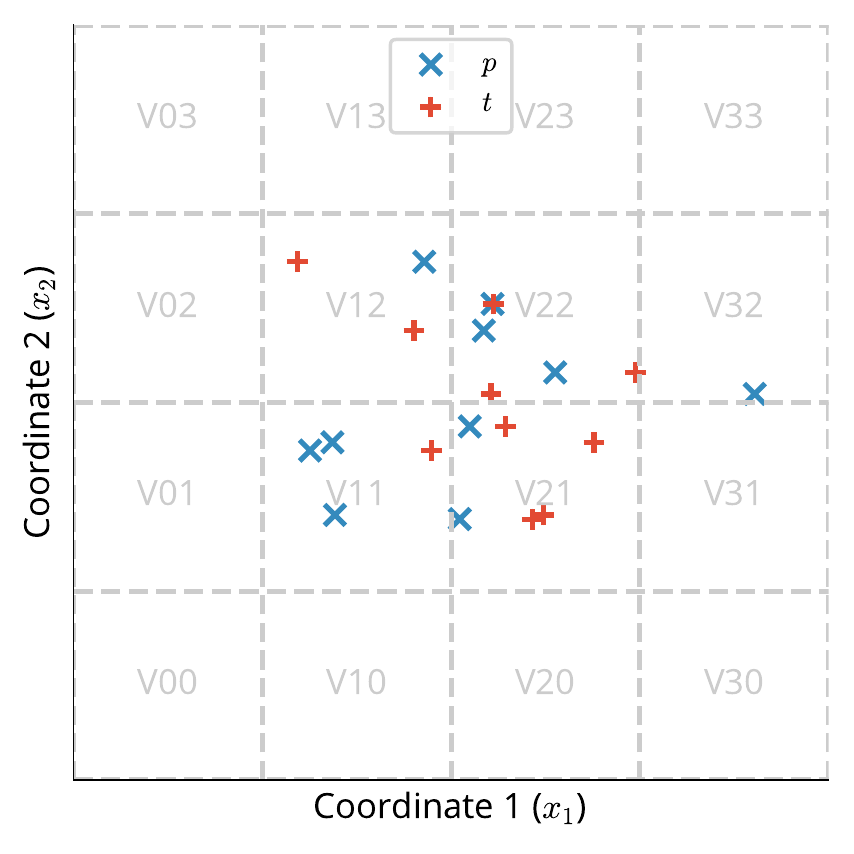}
        \caption{vdKS initially separates the hyperspace into voxels, computing membership within each. For voxels with high membership, ddKS can be performed on that voxel's membership for higher fidelity.}
        \label{fig:vdks_schematic}
    \end{subfigure}
    \begin{subfigure}[b]{\columnwidth}
        \centering
        \includegraphics[width=0.7\textwidth]{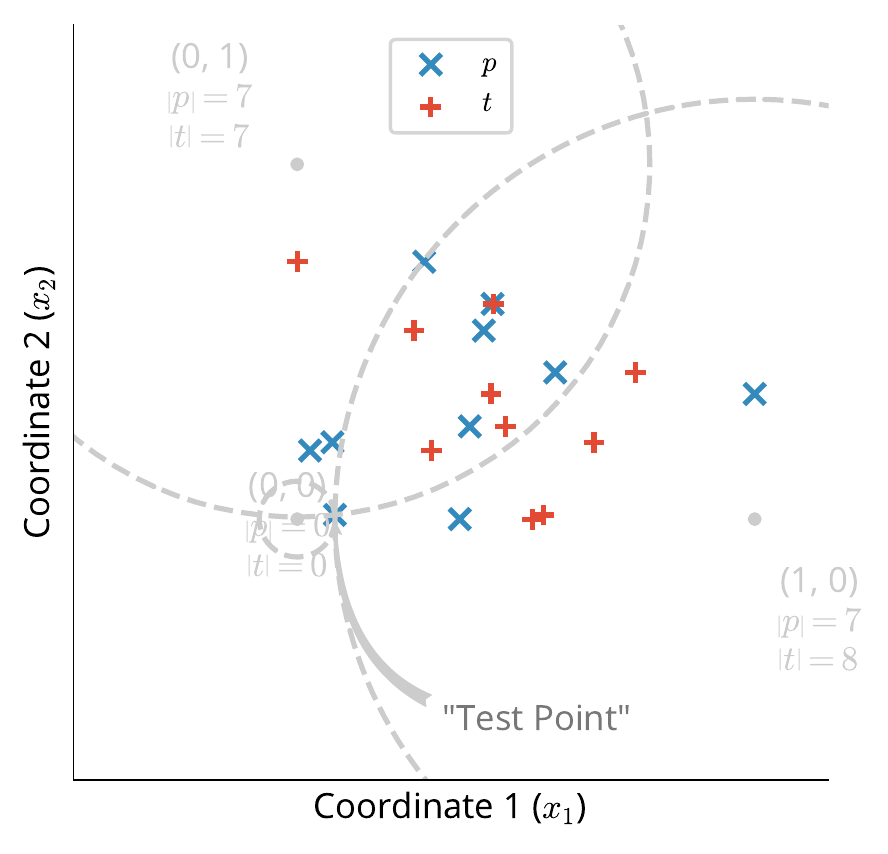}
        \caption{rdKS uses chosen "corners" for comparison, assigning membership for a point to a given quadrant if the euclidean distance between that point and the "corner" is smaller than between the corner and a test point. Note that points can have membership in several quadrants.}
        \label{fig:rdks_schematic}
    \end{subfigure}
    \caption{Schematic Explanations of full \ddKS{} and two accelerated methods, vdKS and rdKS}
    \label{fig:my_label}
\end{figure}

\subsubsection{Voxel based pairwise approximation}

In increasing dimensions, the sparsity of samples falling into many regions of hyperspace will increase exponentially (one of many examples of the curse of dimensionality \cite{Bellman1957}).  Naively using every point in each sample to construct a CDF necessarily spends equal computation on each point in every sample, regardless of the local sparsity of the sample.  Voxel based pairwise approximation \ddKS{} (vdKS) seeks to resolve this issue by dividing space into hypervoxels (hereafter called voxels for simplicity). In each voxel, the membership from each class can be counted for an approximate \ddKS{} distance. In the case of high voxel membership, the full \ddKS{} calculation can be performed for only those points falling in that voxel.  

The algorithm for the voxel based approximation to \ddKS{} (vdKS) is shown in Algorithm~\ref{alg::vdks} and a schematic representation is shown in Figure~\ref{fig:vdks_schematic}.  Without changing the \ddKS{} distance, the sets of $d$ dimensional samples, $\mathcal{P},\mathcal{T}$, are shifted and rescaled to be between 0 and 1 (NormalizeData in Alg \ref{alg::vdks}).  In one pass over all the points in the dataset, each point is assigned one of $k$ equally sized voxels which fill the $d$ dimensional region.  The difference in proportional occupation of $\mathcal{P}, \mathcal{T}$ points is assigned to each voxel.  Calculating the vdKS distance is then finding the max sum of the difference values in each orthant created by splitting space with each non-empty voxel.  SumOrthants returns $2^d$ values, one per orthant.  

\begin{algorithm}
\SetAlgoLined
 \KwResult{Approximates \ddKS{} distance on set of Voxels }
 \KwIn{
 $P$ and $T$ (N,d arrays)\newline
 VoxelPerDim (int)
 }
 \For{id $\in$ [0,1]}{
 \For{index $\in$ [0,VoxelPerDim - 1]} {
 VoxelList[id][index] = 0
 }}
 P,T = NormalizeData($P$,$T$)\\
 FilledVoxels = $\{\}$\\
 D = 0\\
 \For{id,Dataset \textbf{in} enumerate(P,T)}{
     \For{p in Dataset}{
        index = IndexFromPoint(p)
        VoxelList[id][index] += 1\\
        \If{(id,index) not in FilledVoxels}{
            FilledVoxels[(Id,Index)] = 1
        }
    }
    VoxelList[id]/=len(Dataset)}
 Diff = VoxelList[1]-VoxelList[0]\\
 \For{(id,index) in FilledVoxels.keys()}{    
    TmpD = max(SumOrthants(Diff[index]))\\
    D = max(D, TmpD)\\
    }
    return D
 \caption{Voxel d-dimensional KS (vdKS)}
 \label{alg::vdks}
\end{algorithm}

\subsubsection{Radially based pairwise approximation}

\begin{algorithm}
\SetKwFunction{FMain}{GetDFromCornerLists}
\SetKwProg{Fn}{def}{:}{}
  \SetKwProg{Fn}{def}{:}{}

\SetAlgoLined
 \KwResult{Approximates ddKS distance with radial approximation}
 \KwIn{
  $P$ and $T$ (N,d arrays)
 }
 D = 0
 $C$ = IdentifyCorners($P,T$)\\
 \For{$c \in$ ($C)$}{
 PCornerDistances = Sort(GetDistanceFromCorner(P,c))\\
 TCornerDistances = Sort(GetDistanceFromCorner(T,c))\\
 tmpD = GetDFromCornerLists(PCornerDistances,TCornerDistances) \\
 \If{tmpD $>$ D}{
    D = tmpD}
    }
return D
 
\Fn{\FMain{$C1$,$C2$}}{
    numC1 = 0\\
    numC2 = 0\\
    D = 0 \\
    MergeList = []\\
    \While{numC1 $<$ len(C1) or numC2 $<$len(C2)}{
    \uIf{C1[numC1] $<=$ C2[numC2]}{
    numC1 += 1 \\
    }
    \ElseIf{$C2[0] < C1[0]$}{
    numC2 += 1 \\
    }
    tmpD = numC1/len(C1) - numC2/len(C2)
    \If{tmpD $>$ D}{
    D = tmpD}
    }
    return D
}
 \caption{Radial d-dimensional KS (rdKS)}
 \label{alg::rdks}
\end{algorithm}
Like vdKS, the radially based pairwise approximation to ddKS (rdKS) reduces the time complexity with respect to $N$, but it also reduces the time complexity with respect to the dimension, $d$.  A schematic explanation of the rdKS algorithm (Alg: \ref{alg::rdks}) can be found in Figure \ref{fig:rdks_schematic}.  Instead of splitting space via Cartesian axes, we identify $d + 1$ corner points and, for each point, sort the sample points by their distance from each corner.  Instead of comparing orthants, rdKS compares the occupation of spherical volumes of space centered on each corner.  Calculating the occupation of each region is as simple as looking up the point's position on the corresponding radial distance lists.  We note that these spherical regions can overlap allowing for points to be present in multiple regions.   A benefit of using rdKS is replacing the $\mathcal{O}(N^2)$ pairwise calculation with an $\mathcal{O}N\log N$ sorting algorithm.  However, the main benefit comes from reducing the time complexity with regards to dimension.  A naive application of rdKS would create an origin at every corner of the sample space and introduces a $\mathcal{O}(2^d)$ complexity.  Therefore, we approximate the test by selecting only $d + 1$ origins (e.g. for a $3D$ normalised sample space, selecting $(0,0,0), (1,0,0), (0,1,0), (0,0,1)$).

\section{Significance}
\label{sec:sig}
The literature includes several attempts to determine the significance of a given multidimensional KS statistics \cite{Fasano1987, Press1988}, however these require estimates of the covariance matrix of the relations between the two unknown probability distributions. Instead, we present here an analytical calculation of the significance of a ddKS statistic $D$ without computation of the covariance matrix.

In the following we derive a formula for the significance of the two-sample \ddKS{} test.
Here we are concerned with the symmetric ddKS test.   In contrast to the membership matrix of Equation \ref{eq::membership}, we construct membership matrices with respect to both sets ($P,T$) denoted by $\mathbf{M}_{X(P+T)}$.  For brevity, we drop the second index.  The matrix indices of $\mathbf{M}_{P}[i,k]$ ($\mathbf{M}_{T}[i,k]$) range from 0 to $2^d$ and 0 to $N_P (N_T)$ respectively and represent the number of samples which land in a specific orthant. Our objective is to calculate the probability that, under the null hypothesis where the sample sets $P$ and $T$ come from the same distribution, we would see a higher maximum value for the difference in proportional orthant occupation when the orthants are fixed but new samples are generated.  

Each element of $\mathbf{M}_X[i,k]$ can be thought of as the result of $N_X$ binomial trials with success and failure corresponding to the sample landing inside or outside the $i$th of space being split at point $k$ respectively. We note that distribution of numbers of successes into different quadrants given a test point is certainly multinomial, however the number of successes of falling into a certain orthant from \emph{two different samples} is binomial. We are interested in the difference of the latter, not in the former. Thus, each element of the membership matrix follows
\[
\mathbf{M}_{X}\left[i,k\right] \sim Bi\left(N_x, \lambda_{i,k}\right).
\]
Because the PDF for distributions $P$ and $T$ are equivalent under
$H_{0}$, the rate corresponding to each entry in $\mathbf{M}_{P}$
and $\mathbf{M}_{T}$ is equal. The rate $\lambda_{i,k}$ corresponding
to each entry is the total probability density in that orthant, or the integral over the orthant volume $V_{ik}$
\[
\lambda_{i,k}=\int_{V_{ik}}T \,dV.
\]
which is generally unknown. However, it can be estimated because there
exists one set of trials from the desired distribution (those in $T$).
Therefore, we can estimate the rate $\hat{\lambda}_{i,k}$ simply
with the unbiased estimator
\[
\hat{\lambda}_{i,k}=\frac{\mathbf{M}_{T}\left[i,k\right]}{N_T}.
\]
To generalize use to all sample sizes, distributions, and dimensions, we must be capable of predicting a realistic binomial parameter even in the case of zero successes.  In this case, we can use the Bayes estimator with a uniform prior, as is often done for such rare event cases \cite{Razzaghi2002}. In that case,
\[
\hat{\lambda}_{i,k}=\frac{\mathbf{M}_{T}\left[i,k\right]+1}{N_T+2}.
\]
Now, we desire to know
the probability that the absolute difference between the binomial samples
from $P$ and $T$ are as extreme as the value observed. Equivalently, we desire the probability
mass function of the binomially distributed random variables 
\begin{equation}\label{eqn:abs_difference}
\delta=\left|\frac{\mathbf{M}_{P}\left[i,k\right]}{N_{P}}-\frac{\mathbf{M}_{T}\left[i,k\right]}{N_{T}}\right|.
\end{equation}
The desired probability mass function is comprised of the sum of the probability
of all possible ways of achieving a given difference $\delta$. The probability mass
function of a binomial distribution of $n$ successes in $m$ samples with a rate of
$\lambda$ is defined as
\begin{equation}\label{eq:binomial_probability}
p\left(n:m,\lambda\right)=
\left(\begin{array}{c}
m\\
n
\end{array}\right)\lambda^{n}\left(1-\lambda\right)^{m-n}
\end{equation}
Thus, the probability of all possible ways to achieve an absolute normalized
difference $\delta$ given the number of trials in each sample (denoted for sample $X$ as $m_{X}$) and the rate $\lambda$
is
\begin{equation}
p\left(\delta:N_{P},N_{T},\lambda\right)=\sum_{n_{P},n_{T}\in\vec{\Delta}}p_{Bi}\left(n_{P},m_{P},\lambda\right)p_{Bi}\left(n_{T},m_{T},\lambda\right)
\end{equation}
where $\vec{\Delta}$ is the set of possible combinations of numbers of successes
in each sample which would lead to an absolute difference $\delta$.
\begin{multline*}
\vec{\Delta}=\bigg\{ n_{P},n_{T}:\delta=\left|\frac{n_{P}}{N_{P}}-\frac{n_{T}}{N_{T}}\right|,0\leq n_{P}\leq N_{P}, \\
0\leq n_{T}\leq N_{T}\bigg\} 
\end{multline*}
In practice, our estimate of the rate $\hat{\lambda}_{i,k}$ is used
as a substitute for the true rate $\lambda$. We then form
the probability of seeing anything as, or less, extreme than the
difference, $\delta$ observed as
\[
p\left(d\leq \delta:\delta,N_{P},N_{T},\lambda_{i,k}\right)=\sum_{d^{*}=0}^{\delta}p_{\delta}\left(d^{*},N_{P},N_{T},\lambda_{i,k}\right).
\]
The ddKS distance $D$ is the maximum value of all observed $\delta$.
The probability of observing that maximum value $D$ under the null
hypothesis is then the complement of the probability of observing 
this difference $\delta$ or less in every element in $\mathbf{M}$,
therefore the significance $S$ of a ddKS distance $D$ is
\begin{equation}\label{eq:ddks_significance}
S\left(D,N_{P},N_{T},\vec{\lambda}\right)=1.0 - \prod_{i=0}^{2^{d}}\prod_{k=0}^{N}p_{<\delta}\left(D,N_{P},N_{T},\lambda_{i,k}\right).
\end{equation}
This formulation can be easily calculated, although sometimes the binomial coefficient is difficult to compute (especially at large sample sizes).  In the case of large
$m$ and large dimension, $d$, a Poisson approximation can be imposed
(because $\lambda$ can be assumed to be small), replacing $p\left(n:m,\lambda\right)$ (equation~\ref{eq:binomial_probability}) with the probability mass function for a Poisson distribution and propagating through the rest of the derivation. In this way,
equation~\ref{eq:ddks_significance} can be used for any two sample test.

\begin{figure}
    \centering
     \begin{subfigure}[b]{\columnwidth}
         \centering
         \includegraphics[width=\columnwidth]{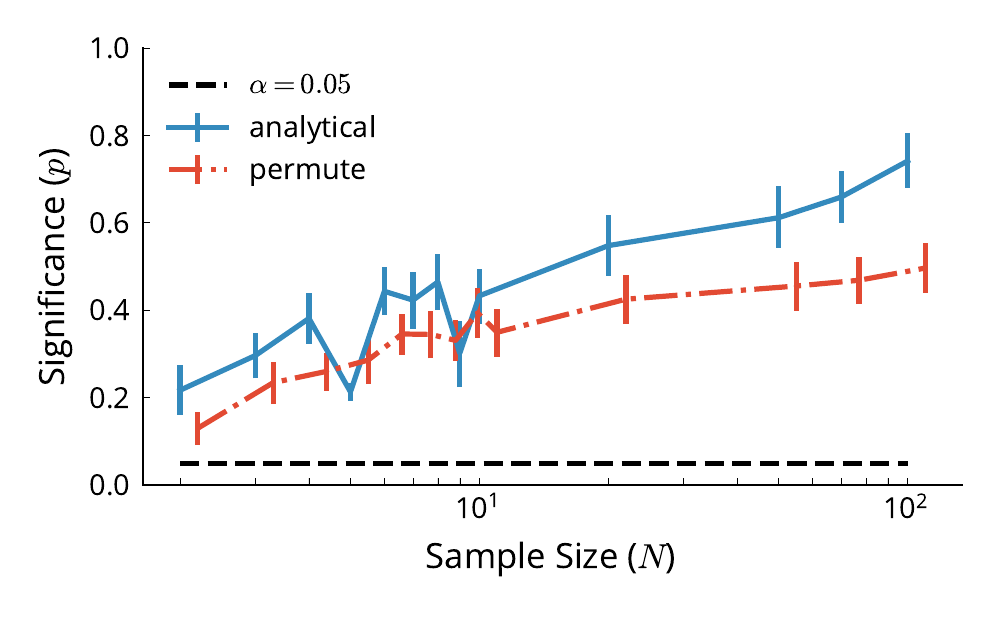}
         \caption{$P$ and $T$ drawn from the same distribution}
         \label{fig:h0_analytical_significance}
     \end{subfigure}
     \vspace{1em}
     \begin{subfigure}[b]{\columnwidth}
         \centering
         \includegraphics[width=\columnwidth]{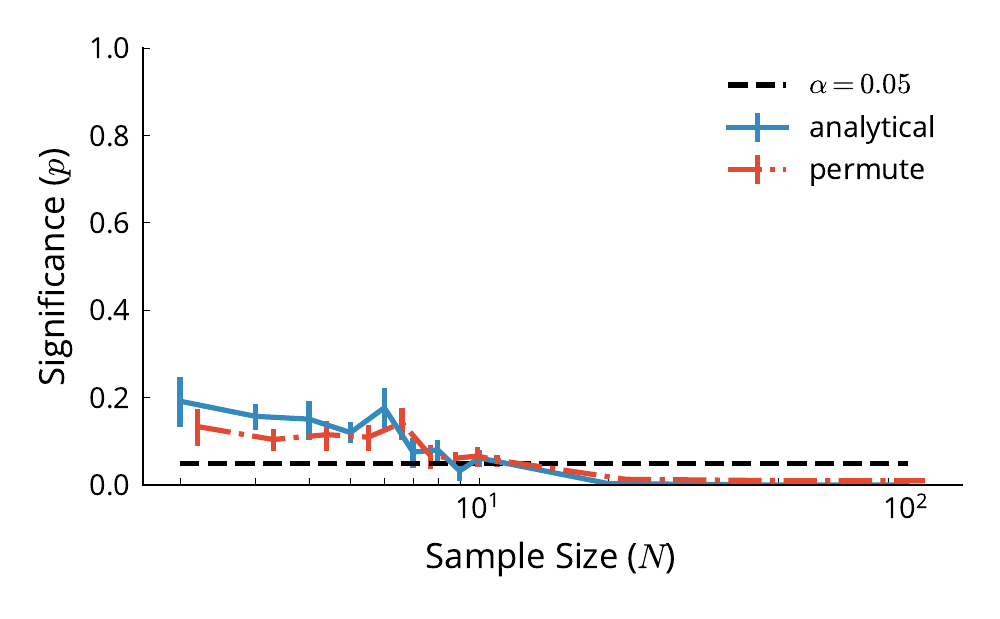}
         \caption{$P$ and $T$ drawn from the different distributions}
         \label{fig:ha_analytical_significance}
     \end{subfigure}
    \caption{Statistical significance with which we can reject $H_0$ with increasing sample size $N$ on the DVU distribution in 3 dimensions, repeated 100 times. Significance using both our analytical formulation and the permutation test are close. When both $P$ and $T$ are from the same distribution, we are not able to reject $H_{0}$ in any case.  When $P$ and $T$ are drawn from different distributions, we are able to consistently reject $H_{0}$ when $N>10$.}
    \label{fig:analytical_significance}
\end{figure}

In figure~\ref{fig:analytical_significance} we show that the analytical significance developed in this section closely matches that estimated using the standard permutation test. When comparing to a sample from the null or alternative distribution, the behavior of the analytical significance closely follows that which was calculated with the permutation method with increasing sample size. In fact, many of the significances are within one standard deviation of each other (we repeated each calculation 100 times).

To illustrate the usefulness of this in a machine learning context, we explore the power of a test on a small sample which has been chosen as it is smaller than the batch size used in much of modern machine learning. We used samples from the DVU dataset (described in Section~\ref{sec:datasets}) to determine the \ddKS{} score and the derivation in this section to determine the significance of that score.  For type I testing, we use two samples from the ``diagonal'' distribution in DVU; for type II testing, we use one sample from the ``diagonal'' and one from the ``uniform'' distribution.  The test for type I error control under $H_{0}$ showed that type I error for the analytical formulation has a 95\% confidence interval of $\left[2.8\%,5.2\%\right]$ for a significance level of $5\%$ when the sample size is 50. The test for the power of the analytical formulation has a 95\% confidence interval of $\left[99.8\%,100.0\%\right]$ for the same significance and sample size  - showing that \ddKS{} has good Type I error control and high statistical power.

Therefore, equation~\ref{eq:ddks_significance} is a general use and accurate
expression for the significance of a given ddKS statistic $D$ under the null
hypothesis, $H_{0}$.

\section{Experiments}
\label{sec:experiments}
In lieu of a functional analysis to compare \ddKS{} to other high dimensional distribution comparison methods, we compare \ddKS{} using multiple datasets.  This method is similar to the comparison shown in \cite{Fasano1987} and \cite{Justel1997}.

\subsection{Datasets}\label{sec:datasets}

\begin{figure*}
    \centering
    \begin{subfigure}[b]{0.30\textwidth}
        \includegraphics[width=\textwidth]{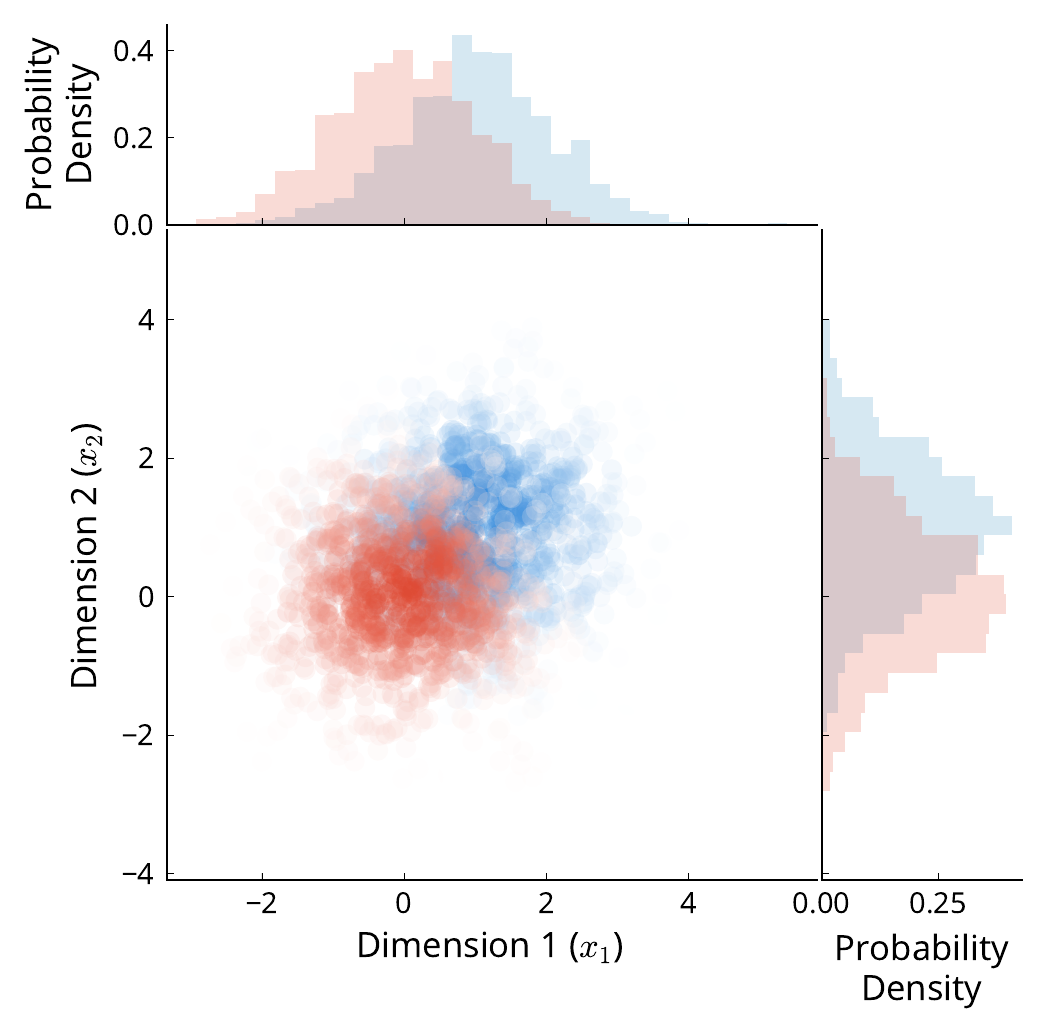}
        \caption{Gaussian varying means}
    \end{subfigure}
    \hfill
    \begin{subfigure}[b]{0.30\textwidth}
        \includegraphics[width=\textwidth]{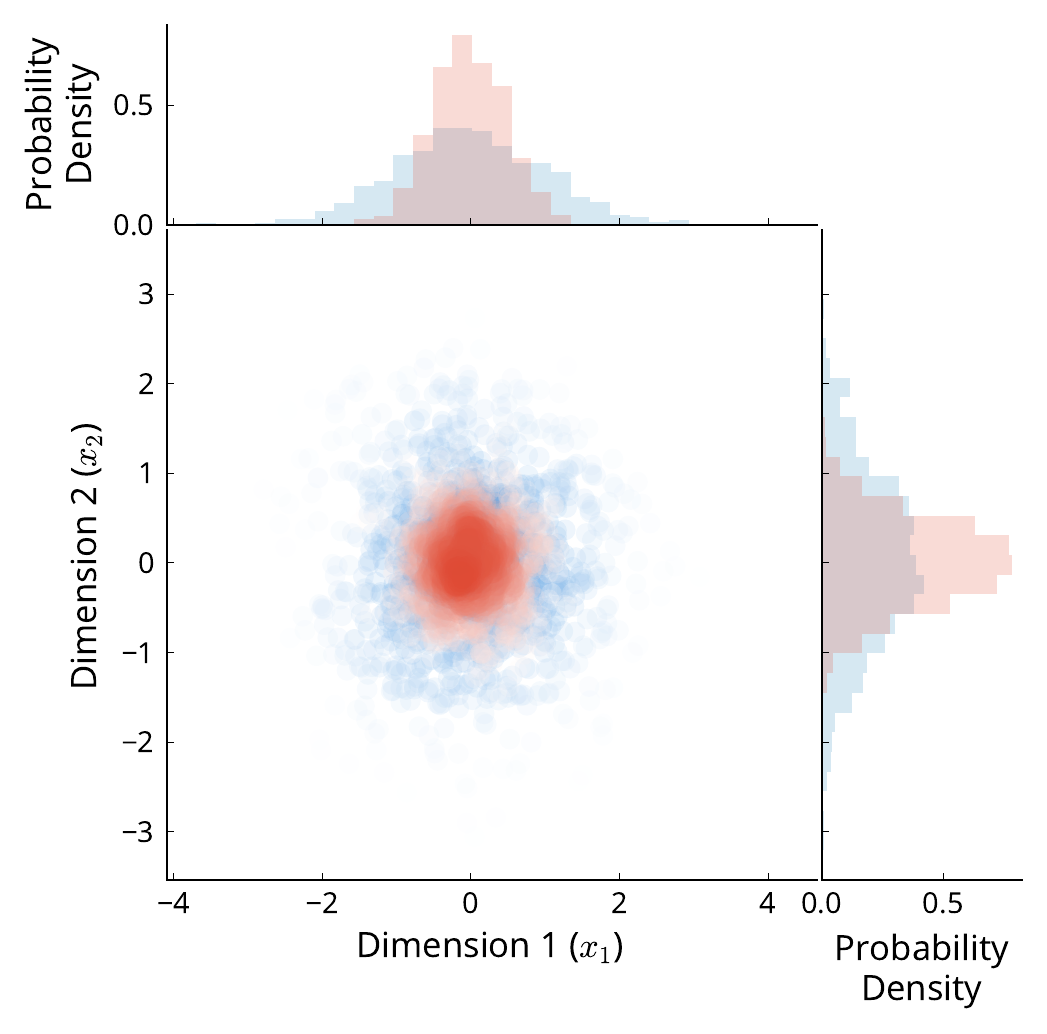}
        \caption{Gaussian varying standard deviations}
    \end{subfigure}
    \hfill
    \begin{subfigure}[b]{0.30\textwidth}
        \includegraphics[width=\textwidth]{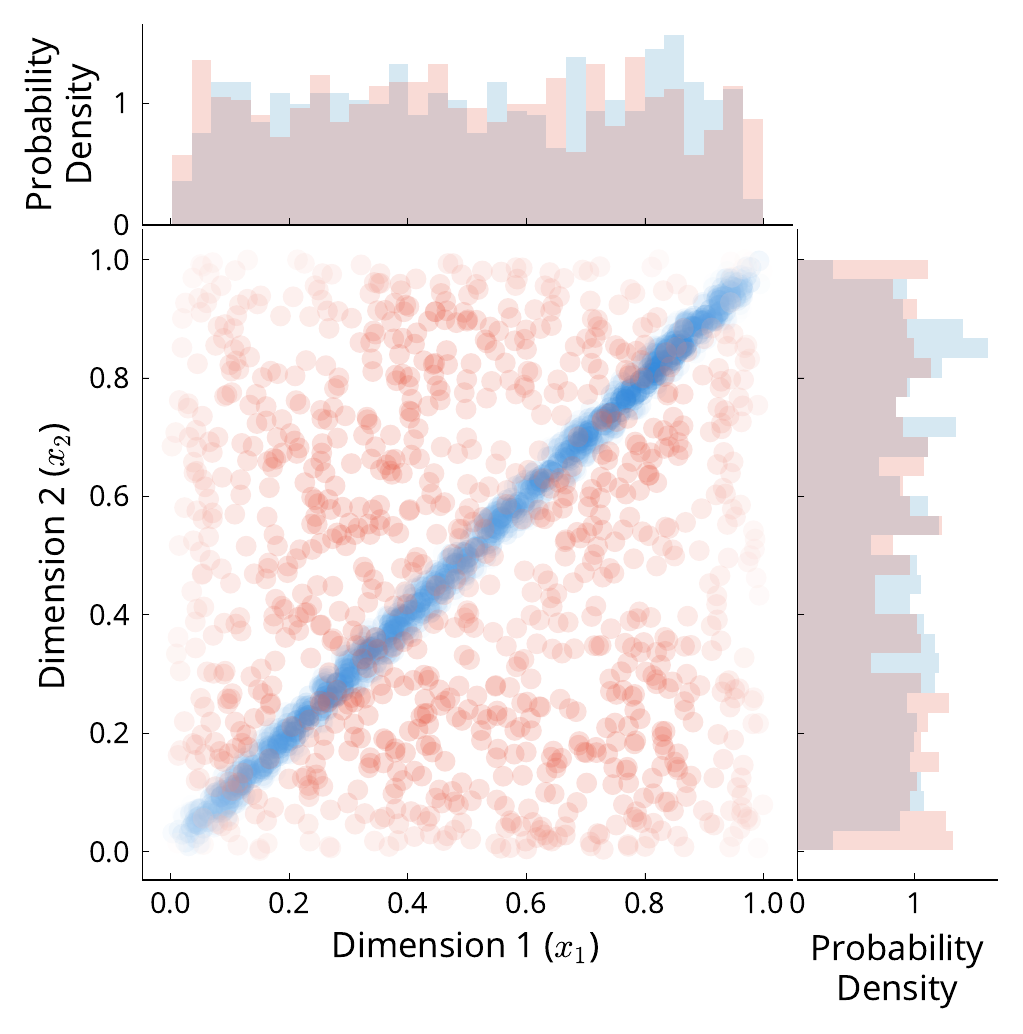}
        \caption{Diagonal versus uniform}
    \end{subfigure}
    \begin{subfigure}[b]{0.30\textwidth}
        \includegraphics[width=\textwidth]{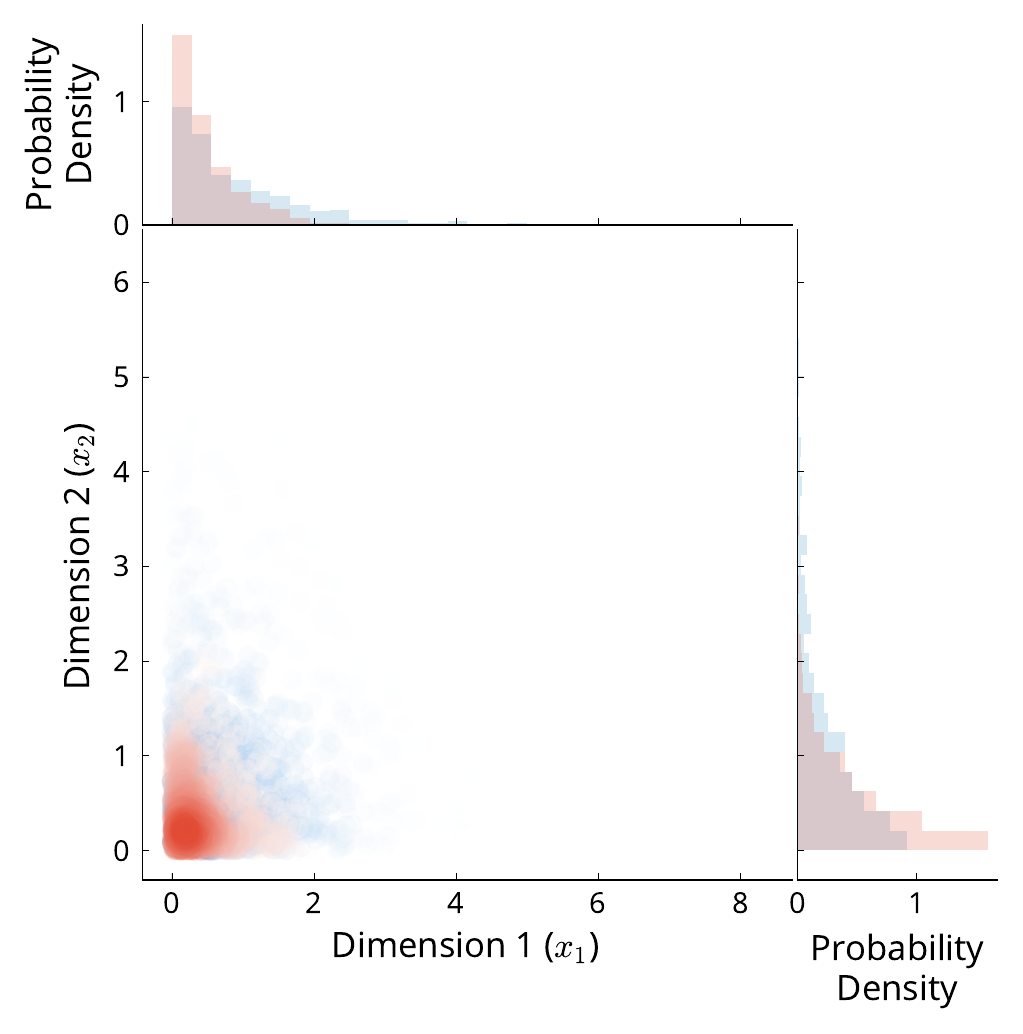}
        \caption{Skewness}
    \end{subfigure}
    \hfill
    \begin{subfigure}[b]{0.30\textwidth}
        \includegraphics[width=\textwidth]{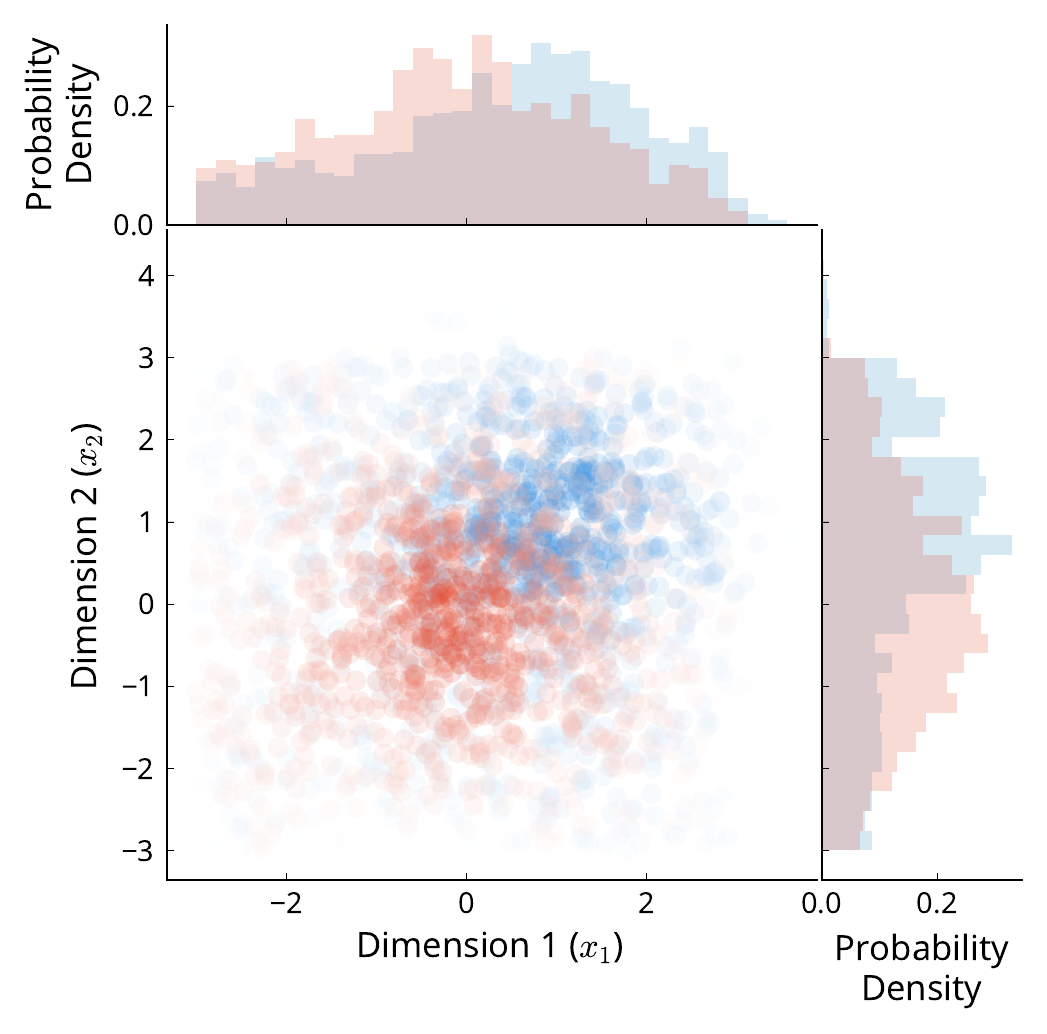}
        \caption{Mixture Model}
    \end{subfigure}
    \hfill
    \begin{subfigure}[b]{0.30\textwidth}
        \includegraphics[width=\textwidth]{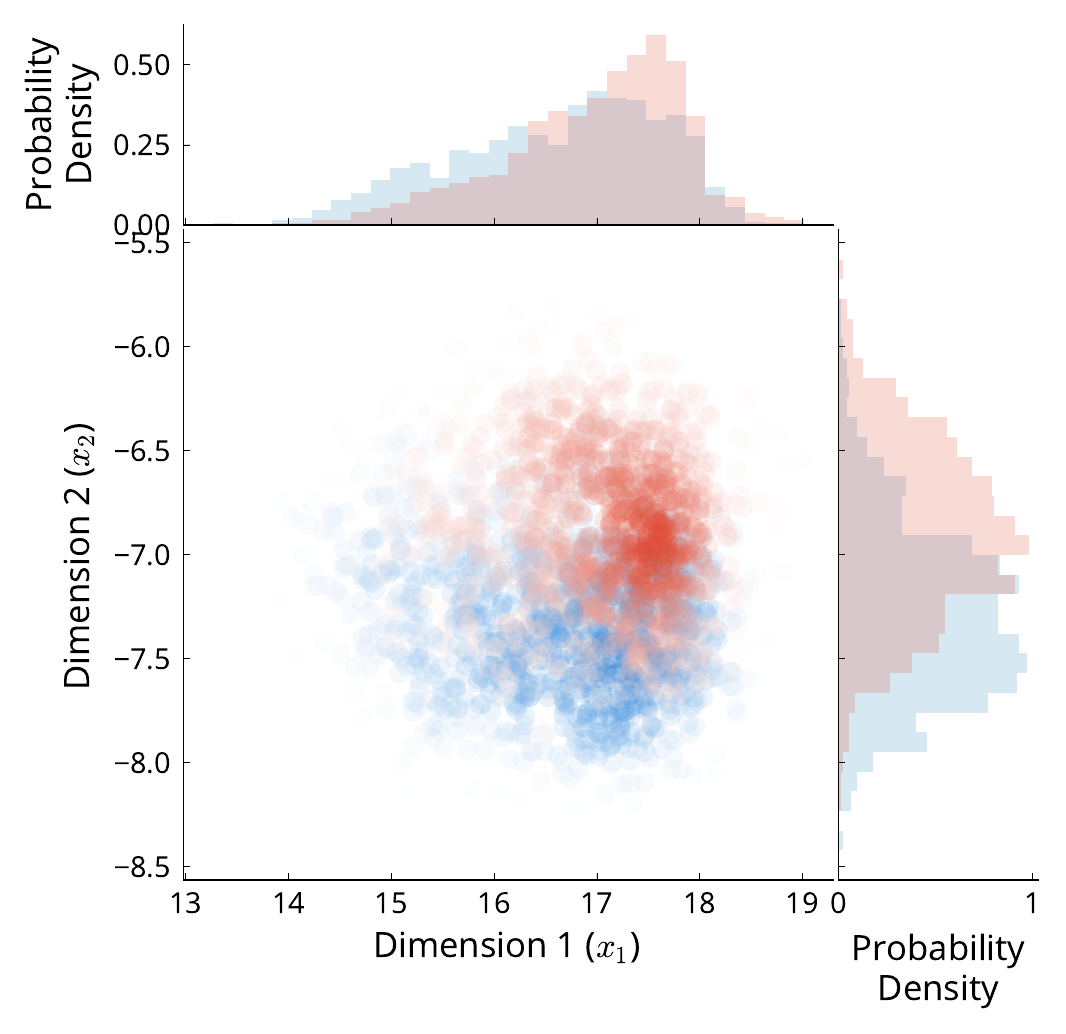}
        \caption{Latent Space}
    \end{subfigure}
    \caption{Illustration of two-dimensional version of all datasets tested}
    \label{fig:datasets}
\end{figure*}

We illustrate the utility of \ddKS{} by exploring its behavior on a variety of datasets, and comparing that behavior to several other standard methods.  We have striven to choose datasets which cover many features encountered in real datasets. Below we describe these datasets, and we illustrate a (two-dimensional) example of each dataset in Figure~\ref{fig:datasets}.

\begin{description}
  \item[\textbf{Gaussian - varying means (GVM)}] \hfill \\ To show that \ddKS{} can discriminate between distribution means in high dimension, we compare two samples from $\mathcal{N}^{3}\left(\mu_{1},\sigma\right)$ and $\mathcal{N}^{3}\left(\mu_{2},\sigma\right)$.
  \item[\textbf{Gaussian - varying standard deviations (GVS)}] \hfill \\ To show that \ddKS{} is sensitive to shape differences, we compare two samples from $\mathcal{N}^{3}\left(\mu,\sigma_{1}\right)$ and $\mathcal{N}^{3}\left(\mu,\sigma_{2}\right)$.
  \item[\textbf{Diagonal versus Uniform (DVU)}] \hfill \\ To show that \ddKS{} can differentiate distributions due to high dimensional covariances, we generate a dataset where the first sample is uniformly sampled along a diagonal from $\left(0, 0, 0\right)$ to $\left(1, 1, 1\right)$, and the second sample is uniformly sampled throughout the space of $\left[0,1\right]^{3}$.
  \item[\textbf{Skewness (Skew)}] \hfill \\ To show that \ddKS{} does not only work with symmetric distributions, we generate two samples from exponential distributions $\text{exp}^{3}\left(\lambda_{1}\right)$ and $\text{exp}^{3}\left(\lambda_{2}\right)$.
  \item[\textbf{Mixture model (MM)}] \hfill \\ To show that \ddKS{} can discriminate small signals in a distribution including noise, we first generate two samples similar to that for "Gaussian - varying means". We then replace a fraction of each sample with those sampled from a uniform distribution from $\left[-3\sigma, 3\sigma\right]$.
  \item[\textbf{Latent space (LS)}] \hfill \\ To illustrate \ddKS{}'s utility in very high dimensions, we take the first $d$ principle components from the latent representation of an image generated by ResNet18 \cite{resnet16} which has been trained on 1000 classes from ImageNet \cite{imagnet09}.  We then compare two samples of these $d$ dimensional representations, comparing two different OpenImages \cite{openimage20} classes (Person and Truck).
\end{description}

\subsection{Methods}

We perform power analysis of \ddKS{} in two ways: we calculate the minimum sample size to correctly reject $H_{0}$ given the default parameters of each dataset, and we calculate the smallest difference in parameters between $p$ and $t$ in each dataset such that $H_{0}$ is rejected with a sample size of 50.  We compare this to several methods from the literature:
\begin{description}
  \item[\textbf{One dimensional Kolmogorov-Smirnov test (OnedKS)}] \hfill \\ We compare our \ddKS{} against one dimensional test statistics by formulating a combined one dimensional KS test in all dimensions.  To do so, we take the maximum of the KS statistic in any dimension.
  \item[\textbf{Hotelling's T2 test (Hotelling-T2)}] \hfill \\ We compare \ddKS{} against a mean-only high dimensional test first published by Hotelling \cite{Hotelling1931}.
  \item[\textbf{Kullback-Leibler Divergence (KLDiv)}] \hfill \\ We compare \ddKS{} to a modern distribution distance, the Kullback-Leibler Divergence \cite{KL51}.  To calculate KLDiv, an estimate of the underlying probability density of each sample is required.  We perform this estimate by taking the $d$-dimensional histogram with constant bin size and bin density defined by Scott's suggestions in \cite{Scott}.
\end{description}

For all methods, even though some methods have analytical formulations of significance, we use the permutation test with 100 permutations to ensure a uniform treatment. To determine the minimum sample size and smallest parameter difference for rejection of $H_{0}$, we use a bisection method to find a sample size/parameter difference which results in a significance within $10^{-3}$ of $\alpha=0.05$.

To investigate the power of \ddKS{} in comparison to other methods at higher dimensions, we perform a similar sample size test with increasing dimensions. While all datasets can be generalized to higher dimensions, the most relevant datasets for high dimensions are the LS, which illustrates the utility of \ddKS{} for modern computer vision methods, and the DVU, which explicitly challenges a two-sample test to discover covariant differences between distributions.

\subsection{Results}\label{sec:results}

\begin{figure}
    \centering
    \includegraphics[width=\columnwidth]{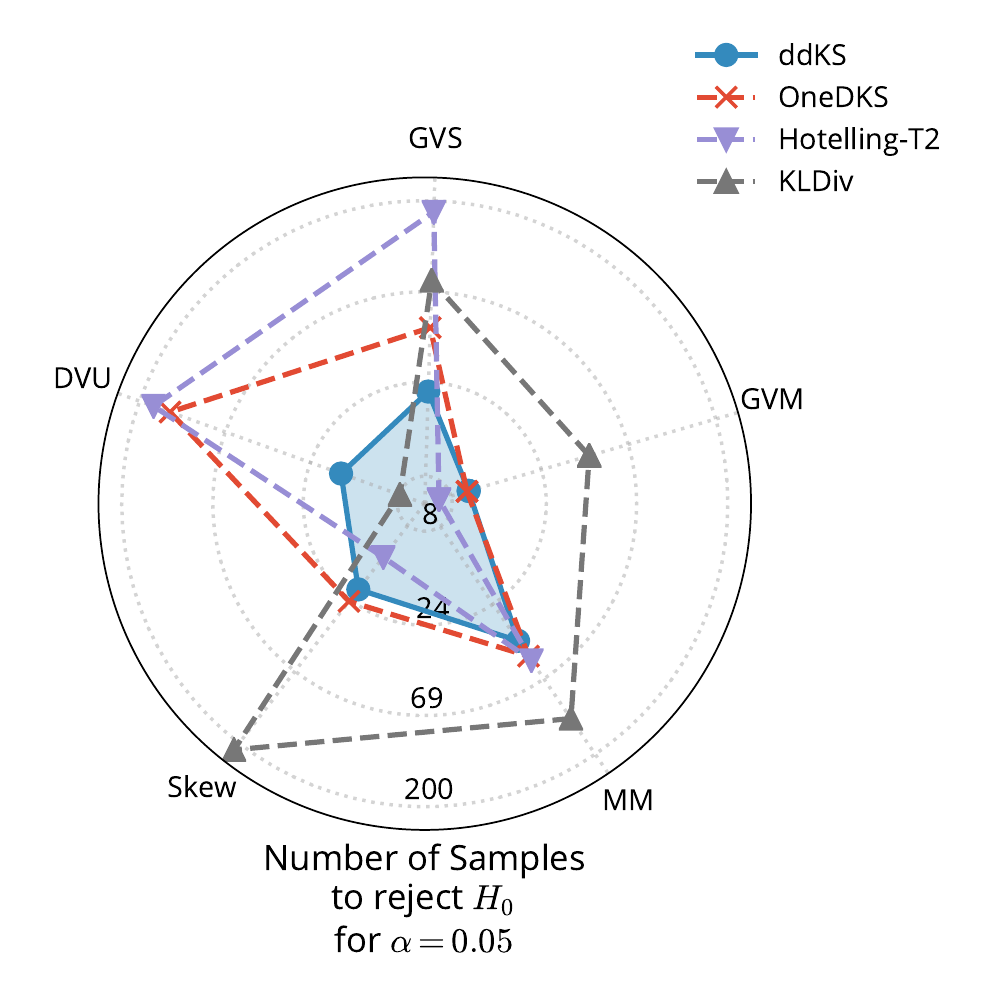}
    \caption{Number of samples required to reject $H_{0}$ for default dataset parameters are plotted radially on a logarithmic scale for each dataset around the circumference. Closer to the center is better. \ddKS{} shows reliably low sample sizes; other methods fail on one or more datasets.}
    \label{fig:n_samples}
\end{figure}

We plot the results of the sample size study in Figure~\ref{fig:n_samples}.  In Figure~\ref{fig:n_samples}, the number of samples required to reject $H_{0}$ is plotted radially on a logarithmic scale for each given dataset.  We notice that, while sometimes \ddKS{} is outperformed, it reliably requires a small number of samples compared to most of the other methods.  We note that, as expected, Hotelling's test required very many samples for two datasets, the GVS and DVU datasets, in which the two samples do not have separate means.  OnedKS had the same difficulty, requiring many samples to differentiate two distributions with the same mean. Interestingly, KLDiv required many samples in all cases except the DVU dataset, on which it outperformed all other methods.

\begin{figure}
    \centering
    \includegraphics[width=\columnwidth]{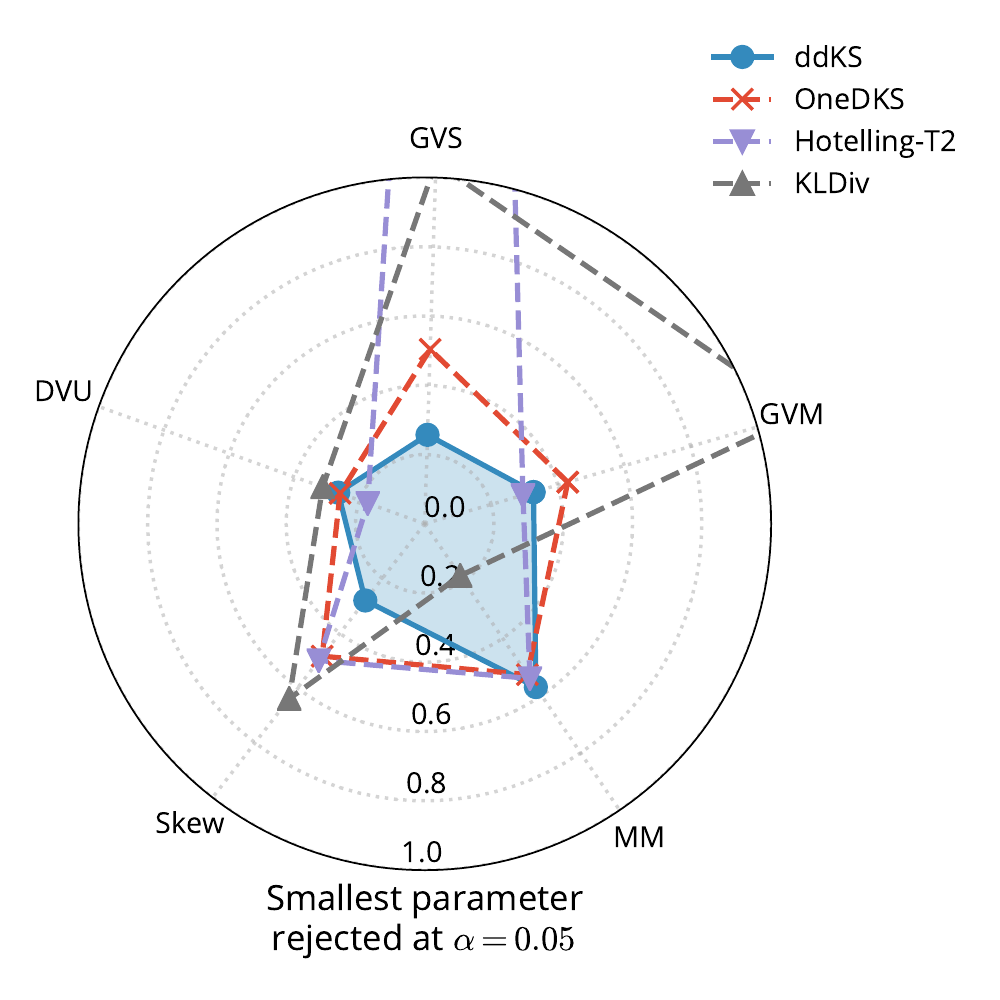}
    \caption{Smallest parameter difference for which a method can reject $H_{0}$ with sample size 50 are plotted radially on a linear scale for each dataset around the circumference. Closer to the center is better. \ddKS{} shows reliably low parameter differences; other methods fail on one or more datasets.}
    \label{fig:shrinkage}
\end{figure}

Figure~\ref{fig:shrinkage} shows, in a similar plot, the smallest parameter for which each method could reject $H_{0}$ with a sample size of 50.  This plot shows the parameter radially on a linear scale for each dataset around the circumference.  Again, \ddKS{} reliably shows high statistical power, with each other method failing on one or more datasets.

\subsection{Alternative Computation Methods}

\begin{figure}
    \centering
    \includegraphics[width=\columnwidth]{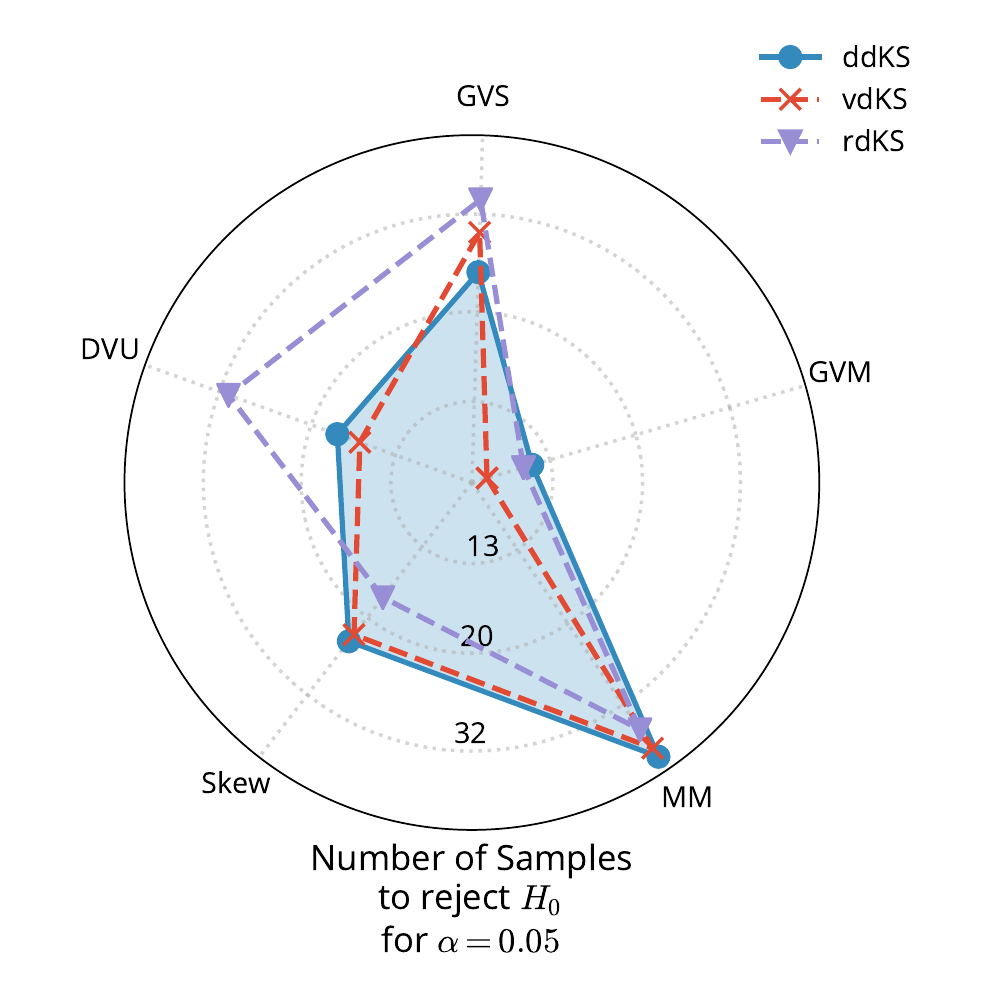}
    \caption{Number of samples required to reject $H_{0}$ for default dataset parameters are plotted radially on a logarithmic scale for each dataset around the circumference. Closer to the center is better. All xdKS methods show similar performance, able to discriminate all datasets investigated with small sample sizes.}
    \label{fig:n_samples_xdks}
\end{figure}

While we have already shown that the alternative computation methods for \ddKS{} are computed with a smaller time complexity, we now show that they correctly approximate the behavior of \ddKS{}.  To do so, we perform a similar power test as before, comparing the number of samples required to reject $H_{0}$ for varying datasets with each different computational method: \ddKS{}, vdKS, and rdKS.  The results are plotted on a radar chart, where the number of samples required to reject $H_{0}$ is plotted as the distance from the center, and the dataset varies around the circumference.  This is shown in Figure~\ref{fig:n_samples_xdks}.

Figure~\ref{fig:n_samples_xdks} shows that the power of each accelerated method is similar on every dataset. Note that the full range of the radial axis is from $\sim$10 to $\sim$50 on this chart, unlike the larger range on Figure~\ref{fig:n_samples}.  By inspection, the most obvious difference is the number of samples required to reject $H_{0}$ for DVU with rdKS, which requires over 30 samples compared to less than 20 samples for vdKS and ddKS.  We attribute this to the overlap between "orthants" in the rdKS computation, although closer inspection is merited in future work.

\begin{figure}
    \centering
    \includegraphics[width=0.95\columnwidth]{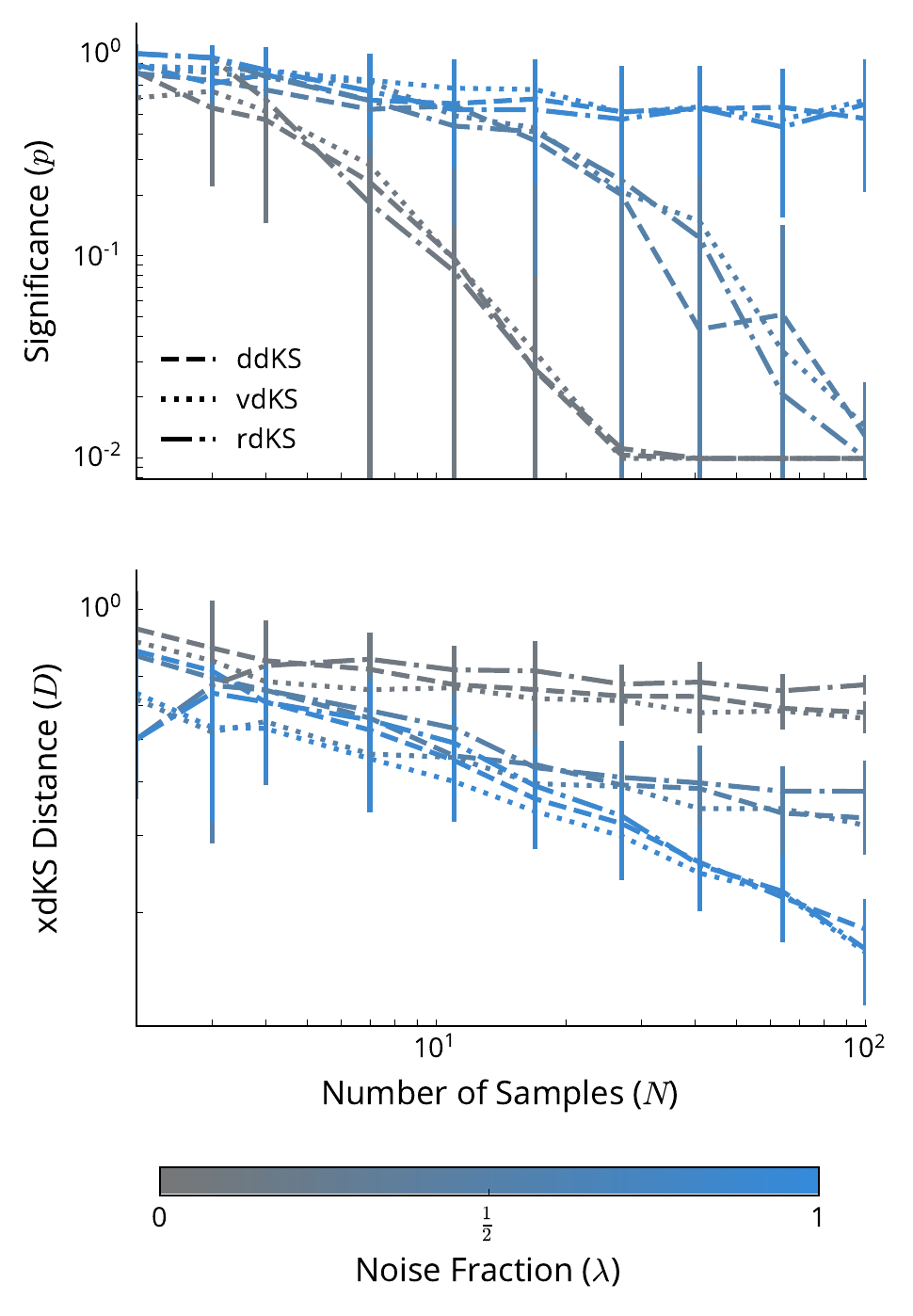}
    \caption{Comparison of xdKS methods raw distances and the significance of that distance for rejecting $H_{0}$ on the MM dataset with given noise fraction and sample size.  All xdKS methods approximate the score to within a few standard deviations, and the signficance even more closely.  Note that the significance decreases to a limit with increasing sample size, whereas the xdKS distance continues decreasing with increasing sample size.}
    \label{fig:xdks}
\end{figure}

We also explore the raw score of each method versus the significance of each score.  To do so, we investigate the MM dataset with increasing rates of noise, recording the raw xdKS score and significance for each sample size.  This is plotted in two panels in Figure~\ref{fig:xdks}.

Broadly, Figure~\ref{fig:xdks} shows that the scores for each accelerated computation are very similar, in fact within one standard deviation (we drew repeated samples and recomputed the score and significance 10 times each).  The same is true for for the significance, reinforcing the claim that each accelerated computation is a good approximation of the full calculation.  One other interesting trend is noted on this chart: that the score continues to decrease even as the significance does not decrease.  We submit that this is a simple sample size effect: as the sample size grows, smaller differences in the empirical CDF are detectable, thus a smaller maximum absolute difference will be of the same significance.

\subsection{Behavior in High Dimensions}

\begin{figure}
    \centering
    \includegraphics[width=0.95\columnwidth]{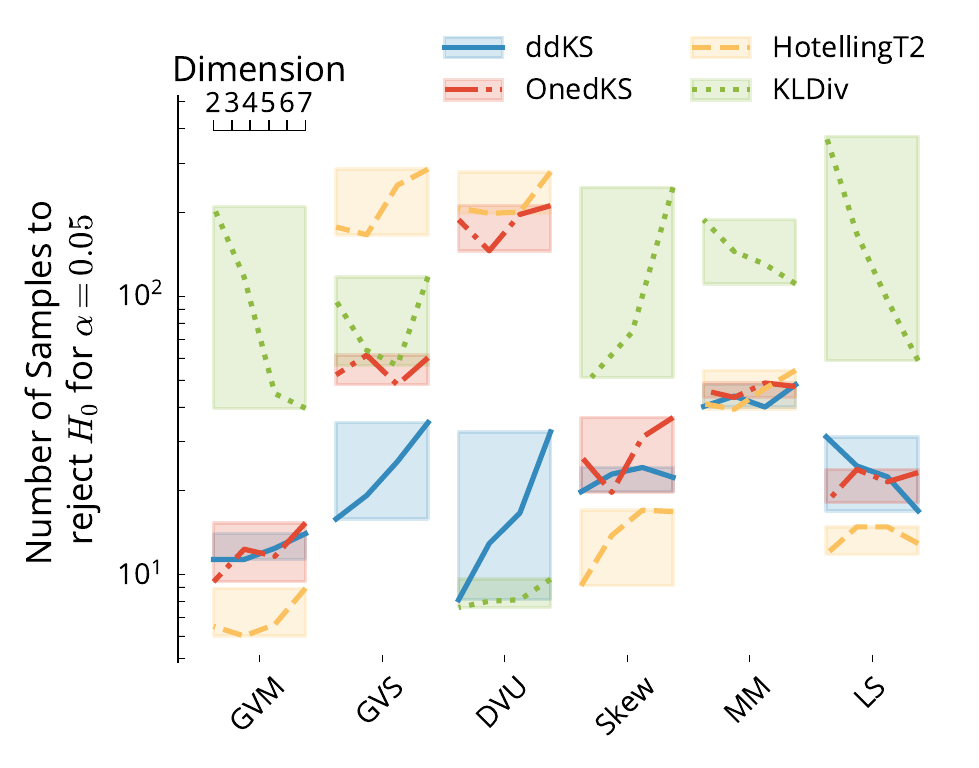}
    \caption{Number of samples to reject $H_{0}$ to a significance of $\alpha=0.05$ for various methods on various datasets at varying dimensions.  The dataset is given on the x-axis and within each dataset the dimension increases across the shaded region.  It can be seen that \ddKS{} performs well on all datasets at all dimensions; Hotelling's T$^{2}$ test and the one dimensional KS test perform poorly unless there is a clear difference in means between the samples, KL Divergence performs poorly except in the case of DVU.}
    \label{fig:ddks_methods_dims}
\end{figure}

This work has striven to extend the KS test to an arbitrary dimension, and we now explore the number of samples required to reject the null hypothesis in increasing dimensions.  We perform the same bisection method to determine the smallest sample size with which $H_{0}$ is rejected for a given dataset, test, and dimension; we repeat this method 10 times to determine the range of possible smallest numbers of samples.  Figure~\ref{fig:ddks_methods_dims} shows the results of this exploration. For each dataset, we plot the results for each test in each tested dimension (2 through 7).  The shaded region shows the overall extents of required number of samples, but the line plot within each shaded region shows the mean value for each method.

In this figure, it can be seen that \ddKS{} reliably requires a modest number of samples to reject $H_{0}$, sometimes performing the best (such as for GVS), but sometimes only performing third best (such as for LS and GVM).  Every other method has several failing, such as Hotelling's T$^2$ failing again on GVS and DVU.  KL Divergence again required many samples except on the DVU dataset.  Another interesting aspect of this figure is the trend with respect to dimension of some methods: ddKS seems to require ever increasing number of samples with increasing dimension for GVS and DVU; KL Divergence requires increasing number of samples for the Skewness dataset, but decreasing number of samples for GVM and LS \footnote{We note that the means of the two distributions in GVM are centered at $\left(0.5, 0.5, \ldots, 0.5\right)$ and thus the distance between distribution means increases with dimension}. Exploration into these trends is planned for our future work.

We also specifically explore the use of rdKS in very high dimensions, and how it could be applied to latent space comparisons in machine learning.  Because we have explicitly formulated rdKS to exhibit linear time complexity with regards to dimension, it is amenable to this use.  We compute the time to compute a single rdKS statistic with 100 points per sample in 1000 dimensions to be approximately 4 seconds.  We also explore the power of that same test (comparing the two distributions that comprise DVU with 100 points per sample in 1000 dimensions).  At the 5\% significance level, rdKS exhibits type I error between 3.7\% and 5.1\% with 95\% confidence.  The power of that test is between 99.6\% and 100.0\% with 95\% confidence, showing that rdKS performs as expected.  A shrinkage study could be performed, however comparison to other test statistics such as \ddKS{} or KLDiv would be computationally intractable. We therefore leave this for future studies.

\section{Conclusions}

In this paper, we have described previous work in the extension of the Kolmogorov-Smirnov test to higher dimensions, and made slight modifications to this extension, naming our result the $d$-dimensional Kolmogorov-Smirnov test (\ddKS{}). These modifications allowed us to make two novel contributions to the literature: the completely analytical calculation of the significance of our \ddKS{} two sample test, and a tensor-primitive based computational method which can be computed very quickly on modern computing hardware.  We admit to the still-significant computational cost of this method, and thus present another novel contribution, two approximate methods (voxel based ddKS and radial-based ddKS) for computation which reduce the time complexity to linear with regards to sample size or linear with regards to dimension.

We explored the behavior of this test against several others, notably Hotelling's T$^2$ test and Kullback-Liebler divergence. This exploration was performed on a broad array of three dimensional datasets, designed to demonstrate power for finding differences in the mean of distributions, finding differences in the shape of distributions, finding only covariant differences, finding differences in non-symmetric distributions, finding differences in distributions corrupted by uniform noise, and finding the differences in latent representations of images from a common computer vision model.  These explorations lead to two main conclusions. The first conclusion is that ddKS performs well on all datasets, whereas each other method has low power on at least one of the explored datasets, and that ddKS is very capable for detecting distribution differences in distributions corrupted by noise. The second conclusion echoes our previous work with ddKS on Cerenkov photon arrival locations and distances \cite{NEURIPS}, which has direct applications in modern high energy physics experiments.

We also explored the behavior of ddKS's accelerated approximations, showing that these are good approximations to \ddKS{} itself, and that they are in fact faster. While \ddKS{} has time complexity of $\mathcal{O}\left(2^{d}N^{2}\right)$, vdKS has time complexity of $\mathcal{O}\left(2^{d}Nk\right)$ and rdKS has time complexity of $\mathcal{O}\left(\left(d+1\right)N\mathrm{log}N\right)$. These time complexities are small enough for use in most physical science applications, and, while still restrictive for the very high dimensional applications present in machine learning fields, represent a significant step towards computationally tractable high dimensional non-parametric test statistic calculation. We proved our dimensional claims by using all methods to analyze higher dimensional distributions, with the results echoing those on the three dimensional distributions.

For a time budget of $1\mathrm{s}$, one can calculate \ddKS{} on two samples with up to $10^{4}$ data points from $\mathrm{R}^{3}$, and, without any assumption about the underlying probability distributions, see extremely high statistical power.  For comparison, using Hotelling's T$^{2}$ test can provide high power and lower computation time, \emph{assuming that the distributions' means differ}, and KLDiv can provide somewhat lower power than \ddKS{} and similar computation time \emph{assuming that the distribution is symmetric}. In higher dimensions, rdKS provides an interface for computation. For a time budget of $5\mathrm{s}$ per computation, rdKS for two samples with $100$ points from $\mathbb{R}^{1000}$ can be computed. For comparison, Hotelling's T$^{2}$ could again provide lower computation time and high power \emph{assuming that the distributions' means differ}, and KLDiv is computationally cost prohibitive.

The combination of all results show that \ddKS{} is a two sample test that can be used to detect mean or shape differences in distributions in dimensions at least up to $1000$, that can be computed quickly on modern computing hardware, and that outperforms other common methods in many cases, specifically the case of distributions corrupted with noise. We believe this will have impact into the validation, and perhaps training, of results from modern data scientific methods such as the latent representations of images; we also believe it will be useful to compare surrogate models to their desired distribution and for signal analysis in high dimensional data in the physical sciences.

\section{Acknowledgements}

We gratefully acknowledge the United States Department of Energy Office of Science's funding and support on this work. James Kahn's work is supported by the Helmholtz Association Initiative and Networking Fund under the Helmholtz AI platform grant. Isabel Haide's work was supported by the Federal Ministry of Education and Research of Germany (BMBF). We acknowledge the Belle2 collaboration for inspiration and permission to work on data scientific problems pertaining to the experiment.  We also gratefully thank Markus G\"{o}tz and Panos Stinis for careful reading and insightful comments.

\bibliographystyle{ieeetr}
\bibliography{references}

\begin{appendices}

\subsection{Proofs} \label{sec:proofs}

\subsubsection{Identity of indiscernibles}

To prove the identity of indiscernibles, we must prove that $D=0$ when $\mathbf{P}=\mathbf{T}$. Through the definition of $\mathbf{C}_{P}$ and $\mathbf{C}_{T}$, these are also equal, thus:
$$\mathbf{G}_{P,P} = \mathbf{G}_{P,T} = \mathbf{G}_{T,P} = \mathbf{G}_{T,T}$$
and
$$\mathbf{M}_{P,P} = \mathbf{M}_{P,T} = \mathbf{M}_{T,P} = \mathbf{M}_{T,T}$$
which ensures that both
$$\left|\mathbf{M}_{P,P} - \mathbf{M}_{T,P}\right| = 0,\;\left|\mathbf{M}_{P,T} - \mathbf{M}_{T,T}\right| = 0\; \therefore D = 0 $$

\subsubsection{Symmetry}

We calculate $D$ using both samples as test points, which makes the test statistic symmetric.  This can be easily proven by the commutative property of the $\mathrm{max}$ operator.  We take the maximum of the concatenation of the two sets $\left|\mathbf{M}_{P,P} - \mathbf{M}_{T,P}\right|$ and $\left|\mathbf{M}_{P,T} - \mathbf{M}_{T,T}\right|$. It is clear, by construction, that if $\mathbf{P}$ and $\mathbf{T}$ are exchanged, then so are $\mathbf{C}_{P,T}$ with $\mathbf{C}_{T,P}$ and $\mathbf{C}_{P,P}$ with $\mathbf{C}_{T,T}$. A similar exchange happens for $\mathbf{G}$ and $\mathbf{M}$.  This leads to the exchange of the two sets $\left|\mathbf{M}_{P,P} - \mathbf{M}_{T,P}\right|$ and $\left|\mathbf{M}_{P,T} - \mathbf{M}_{T,T}\right|$, but the maximum of the concatenation of these two sets is commutative, so this has no effect on the maximum value, $D$, which is the \ddKS{} distance.

\subsubsection{Subadditivity}

We first note that the maximum absolute value of any set is its $l$-$\infty$ norm, which has a subadditivity property of its own; thus \ddKS{} is subadditive.  We provide the following proof for further elucidation.

We prove subadditivity following the method for the exposition of subadditivity for the euclidean distance from \cite[pp. 16-17, 30]{Rudin1976}.  The ddKS distance is defined as
\begin{equation}
    d\left(\mathbf{X},\mathbf{Y}\right) \equiv \max\left[ \left| \mathbf{X} - \mathbf{Y} \right| \right]
\end{equation}
where $\mathbf{X}$, $\mathbf{Y}$ are $d$-dimensional cumulative density functions, that is all elements of $\mathbf{X}$ and $\mathbf{Y}$ are between zero and one, and the sum of all elements in either $\mathbf{X}$ or $\mathbf{Y}$ is one. The $\max$ function operates over all elements in the tensor, the subtraction operator is element-wise, and the pipe operator is the absolute value, operating in the usual way.  We can prove that the $\max\left[\left|\;\right|\right]$ operator shows subadditivity for any tensor whose elements are in $\mathbb{R}^{d}$ by first proving that 
\begin{equation}
    \max\left[ \left| \mathbf{\Lambda} + \mathbf{\Psi} \right| \right] \leq \max\left[ \left| \mathbf{\Lambda} \right| \right] + \max\left[ \left| \mathbf{\Psi} \right| \right]
\end{equation}
To do, so we square the left and right side, and use the Schwartz inequality, seeing that
\begin{align*}
\left(\max\left[\left|\mathbf{\Lambda}+\mathbf{\Psi}\right|\right]\right)^{2}\leq & \max\left[\left|\mathbf{\Lambda}\right|\right]^{2}+2\max\left[\left|\mathbf{\Lambda}\right|\right]\max\left[\left|\mathbf{\Psi}\right|\right]\\
 & +\max\left[\left|\mathbf{\Psi}\right|\right]^{2}\;\therefore\\
\left(\max\left[\left|\mathbf{\Lambda}+\mathbf{\Psi}\right|\right]\right)^{2} & \leq\left(\max\left[\left|\mathbf{\Lambda}\right|\right]+\max\left[\left|\mathbf{\Psi}\right|\right]\right)^{2}\quad\therefore\\
\max\left[\left|\mathbf{\Lambda}+\mathbf{\Psi}\right|\right] & \leq\max\left[\left|\mathbf{\Lambda}\right|\right]+\max\left[\left|\mathbf{\Psi}\right|\right]
\end{align*}
Then, we can prove the subadditivity property for the ndKS distance by replacing $\mathbf{\Lambda}$ with $\mathbf{X} - \mathbf{Z}$ and replacing $\mathbf{\Psi}$ with $\mathbf{Z-Y}$
\begin{align*}
\left(\max\left[\left|\left(\mathbf{X}-\mathbf{Z}\right)+\left(\mathbf{Z}-\mathbf{Y}\right)\right|\right]\right)^{2} & \leq\max\left[\left|\left(\mathbf{X}-\mathbf{Z}\right)\right|\right]^{2}\\
 & +2\max\left[\left|\left(\mathbf{X}-\mathbf{Z}\right)\right|\right]\max\left[\left|\left(\mathbf{Z}-\mathbf{Y}\right)\right|\right]\\
 & +\max\left[\left|\left(\mathbf{Z}-\mathbf{Y}\right)\right|\right]^{2}\;\therefore\\
\left(\max\left[\left|\left(\mathbf{X}-\mathbf{Z}\right)+\left(\mathbf{Z}-\mathbf{Y}\right)\right|\right]\right)^{2} & \leq\left(\max\left[\left|\left(\mathbf{X}-\mathbf{Z}\right)\right|\right]\right. \\
 & +\left.\max\left[\left|\left(\mathbf{Z}-\mathbf{Y}\right)\right|\right]\right)^{2}\quad\therefore\\
\max\left[\left|\left(\mathbf{X}-\mathbf{Z}\right)+\left(\mathbf{Z}-\mathbf{Y}\right)\right|\right] & \leq\max\left[\left|\left(\mathbf{X}-\mathbf{Z}\right)\right|\right]\\
 & +\max\left[\left|\left(\mathbf{Z}-\mathbf{Y}\right)\right|\right]\quad\therefore\\
\max\left[\left|\mathbf{X}-\mathbf{Y}\right|\right] & \leq\max\left[\left|\left(\mathbf{X}-\mathbf{Z}\right)\right|\right]\\
 & +\max\left[\left|\left(\mathbf{Z}-\mathbf{Y}\right)\right|\right]
\end{align*}
Therefore, the ddKS exhibits subadditivity.

\end{appendices}
\end{document}